\documentclass[aps,twocolumn,showpacs,superscriptaddress,nofootinbib]{revtex4}

\usepackage{graphicx}
\usepackage{latexsym}

\usepackage{color}

\begin{document}

\title{Two phase picture in driven polymer translocation}

\author
{Takuya Saito\footnote{
Present address:
Fukui Institute for Fundamental Chemistry, Kyoto University, Kyoto 606-8103, Japan,
\\
$\ast$Electric mail:saito@fukui.kyoto-u.ac.jp}
}
\affiliation{Department of Physics, Kyushu University 33, Fukuoka 812-8581, Japan}

\author{Takahiro Sakaue}
\email[Electric mail:]{sakaue@phys.kyushu-u.ac.jp}
\affiliation{Department of Physics, Kyushu University 33, Fukuoka 812-8581, Japan}
\affiliation{PRESTO, Japan Science and Technology Agency (JST), 4-1-8 Honcho Kawaguchi, Saitama 332-0012, Japan}

\def\Vec#1{\mbox{\boldmath $#1$}}
\def\degC{\kern-.2em\r{}\kern-.3em C}

\def\SimIneA{\hspace{0.3em}\raisebox{0.4ex}{$<$}\hspace{-0.75em}\raisebox{-.7ex}{$\sim$}\hspace{0.3em}} 

\def\SimIneB{\hspace{0.3em}\raisebox{0.4ex}{$>$}\hspace{-0.75em}\raisebox{-.7ex}{$\sim$}\hspace{0.3em}} 

\def\degC{\kern-.2em\r{}\kern-.3em C}

\def\gsim{\hspace{0.3em}\raisebox{0.5ex}{$>$}\hspace{-0.75em}\raisebox{-.7ex}{$\sim$}\hspace{0.3em}} 

\def\lsim{\hspace{0.3em}\raisebox{0.5ex}{$<$}\hspace{-0.75em}\raisebox{-.7ex}{$\sim$}\hspace{0.3em}} 

\date{\today}

\begin{abstract}
Two phase picture is a simple and effective methodology to capture the nonequilibrium dynamics of polymer associated with tension propagation.
When applying it to the driven translocation process, there is a point to be noted, as briefly discussed in our recent article [Phys. Rev. E \textbf{85}, 061803 (2012)].
In this article, we address this issue in detail and modify our previous prediction [Euro. Phys. J. E {\bf 34}, 135 (2011)] by adopting an alternative steady-state ansatz. 
The modified scaling prediction turns out to be the same as that of the iso-flux model recently proposed by Rowghanian and Grosberg [J. Phys. Chem. B {\bf 115}, 14127-14135 (2011)].
\end{abstract}


\def\degC{\kern-.2em\r{}\kern-.3em C}

\maketitle

\def\Vec#1{\mbox{\boldmath $#1$}}
\def\degC{\kern-.2em\r{}\kern-.3em C}

\def\SimIneA{\hspace{0.3em}\raisebox{0.4ex}{$<$}\hspace{-0.75em}\raisebox{-.7ex}{$\sim$}\hspace{0.3em}} 

\def\SimIneB{\hspace{0.3em}\raisebox{0.4ex}{$>$}\hspace{-0.75em}\raisebox{-.7ex}{$\sim$}\hspace{0.3em}} 

\def\degC{\kern-.2em\r{}\kern-.3em C}

\def\gsim{\hspace{0.3em}\raisebox{0.5ex}{$>$}\hspace{-0.75em}\raisebox{-.7ex}{$\sim$}\hspace{0.3em}} 

\def\lsim{\hspace{0.3em}\raisebox{0.5ex}{$<$}\hspace{-0.75em}\raisebox{-.7ex}{$\sim$}\hspace{0.3em}}

Tension-propagation mechanism is inherent in the dynamical response of polymers to local stimuli.
Sometime ago, it has been proposed as an essential ingredient governing the driven translocation process~\cite{PRE_Sakaue_2007}. Such a view seems to have been well validated in recent studies~\cite{Ikonen_Sung_2011}.

In ref.~\cite{Saito_EPJE,Saito_process}, we analyzed in detail the translocation dynamics based on two phase formalism, in which the portion of the translocation polymer in {\it cis} side is pictured to be composed of the moving and the yet quiescent equilibrium domains.
The point of the formalism is to focus on the dynamics of the domain boundary, which is a gross variable describing how the driving force at the pore site is transmitted along the chain backbone.
For such a coarse-graining to be meaningful, other degrees of freedom need to be properly eliminated, the effect of which should be reflected in the behaviors of the gross variable.
Most naturally, the moving domain is assumed to be in a steady-state with a characteristic velocity $V(t)$. Then, the steady-state dynamical friction/extension laws are coupled to the continuum equation describing the conservation of the mass.
In our earlier article~\cite{Saito_EPJE}, we set $V(t) \equiv v_R(t)$, where $v_R(t)$ represents the segment velocity around the rear end of the moving domain.
This steady-state ansatz, however, turned to be not compatible with the mass conservation relation inherent in the translocation process.
This point was shortly mentioned in the appendix of our article~\cite{Saito_process}.
To make the point clearer, we present a detailed account in this
article, 
where our previous result~\cite{Saito_EPJE} is modified
by adopting an alternative steady-state ansatz suggested in refs.~\cite{Saito_process,Arxiv_Dubbeldam_2011}.
We then point out that the alternative ansatz is intimately related with the iso-flux model~\cite{JPCB_Rowghanian_Grosberg_2010}.

\begin{figure}
\begin{center}
\includegraphics{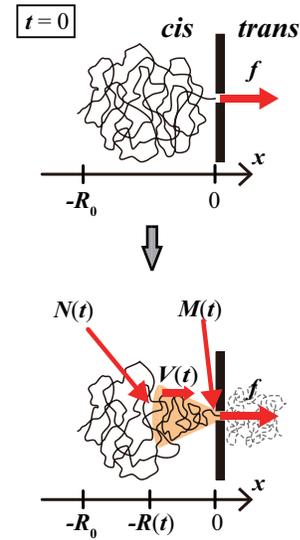}
      \caption{
	  (Color Online) Illustration of the driven polymer translocation process.
	  }
\label{fig1}
\end{center}
\end{figure}

{\it Basic equations}---
Let us consider the driven translocation process as schematically depicted in Fig.~1.
A polymeric chain is initially dissolved in the solution ($x \leq 0$), taking a random coil conformation with the overall size $R_0 \simeq a N_0^\nu$, where $N_0$ is the total segment number, $a$ is the segment length, and $\nu$ is the Flory exponent.
As those in ref.~\cite{Saito_EPJE}, the length, force and time are made dimensionless by the units $a$, $k_{\rm B}T/a$ and $\eta a^3/k_{\rm B}T$, respectively ($k_{\rm B}T$ is the thermal energy; $\eta$ is the viscosity of the solvent).

There is a thin wall at $x=0$ with a small pore, where the driving force with the constant magnitude $f$ is locally exerted in $x$-direction from the {\it cis} to the {\it trans} side. 
One chain end is sucked at time $t=0$. The translocation process, then, proceeds with the tension propagation along the chain.
Segments are numbered from the first sucked end to the other end ($N_0$th segment).
According to the so-called two phase formalism, the dynamics of driven translocation is described as the growth process of the moving tensed domain, which is specified by the segment $N(t)$ at the end of the moving domain, the segment $M(t)$ at the pore, the size $R(t)$ and the representative (or average) velocity $V(t)$ of the moving domain at time $t$ (See Fig.~1). 
These can be determined by the following set of equations:
\begin{eqnarray}
\frac{{\rm d}N(t)}{{\rm d}t}-\frac{{\rm d}M(t)}{{\rm d}t}&=& j_R(t) - j_0(t) + \sigma_R(t) \frac{{\rm d}R}{{\rm d}t}, 
\label{eq:1}
\end{eqnarray}
\begin{eqnarray}
V(t)R(t) &\simeq& f^{p_z}
\label{eq:2}
\end{eqnarray}
\begin{eqnarray}
N(t)-M(t) &\simeq& R f^{-p_\nu} \ (\simeq R \sigma_0)
\label{eq:3}
\end{eqnarray}
\begin{eqnarray}
N(t)^{\nu} &\simeq& R(t)
\label{eq:4}
\end{eqnarray}
where $j_R(t) = \sigma_R(t) v_R(t)$ and $j_0(t) = \sigma_0(t) v_0 (t)$ are the segment fluxes at the moving domain boundary and at the pore, respectively.
As explained in ref.~\cite{Saito_EPJE}, eq.~(\ref{eq:1}) is the continuum equation, eqs.~(\ref{eq:2}) and~(\ref{eq:3}) are the dynamical equations of state describing the steady-state relation among velocity-extension-force (eq.~(\ref{eq:2})) and mass-extension-force (eq.~(\ref{eq:3})) for a dragged polymer, where $p_z = z-2$ and $p_\nu=(1-\nu)/\nu$ in trumpet regime $f_\sharp \leq f \le f^*$, while  $p_z = 1$ and $p_\nu=0$ for stronger force $f^* \le f$~\footnote{For the definition of the characteristic force $f_\sharp \simeq N_0^{-\nu}$, $f^* \simeq 1$ and $f^{**} \simeq N_0^\nu$, see ref.~\cite{Saito_EPJE}.
Strong force regimes are further classified as the stem-flower ($f^*<f<f^{**}$) and the strong-stretching ($f^{**}<f$).}.
Note that, while the segment line density at the pore $\sigma_0 \simeq f^{-p_\nu}$ is given by the exponent related to equilibrium (static) situation, eq.~(\ref{eq:2}) reflects the dissipation mechanism through the exponent $p_z$, which contains the dynamical exponent $z$ characterizing the dynamics of ``critical fluctuation".

{\it Ansatz (i)}---
In ref.~\cite{Saito_EPJE}, we eliminate the pore flux
\begin{eqnarray}
\frac{{\rm d}M(t)}{{\rm d}t} =j_0(t) =  \sigma_0 v_0(t)
\label{eq:5}
\end{eqnarray}
in eq.~(\ref{eq:1}) and made the steady-state ansatz for the moving domain by setting $v_R(t) = V(t)$, which leads to 
\begin{eqnarray}
\sigma_R(t) \left( V(t) +  \frac{{\rm d}R(t)}{{\rm d}t} \right)= \frac{{\rm d}N(t)}{{\rm d}t}. 
\label{eq:6}
\end{eqnarray}
Combining this with eqs.~(\ref{eq:2}),~(\ref{eq:4}) and the segment line density at the boundary $\sigma_R(t) \simeq v_R(t)^{-q} \simeq V(t)^{-q}$, where $q=(1-\nu)/[(z-1)\nu]$ for $f_\sharp \le f \le f^{**}$ and $q=0$ for larger force, one obtains the tension-propagation law
\begin{eqnarray}
R(t) \simeq (t f^{\beta})^{\gamma}
\label{eq:7}
\end{eqnarray}
with $\beta=p_z(1-q)$ and $\gamma = \nu/(1+\nu-\nu q)$, thus, the scaling formula for propagation time is 
\begin{eqnarray}
\tau_{\rm p} \simeq N_0^{\alpha}f^{-\beta}
\label{eq:8}
\end{eqnarray}
with $\alpha = \nu/\gamma$.

{\it Ansatz (ii)}---
An alternative steady-state ansatz is to set $v_0(t) = V(t)$ as was done in ref.~\cite{Saito_process,Arxiv_Dubbeldam_2011}. Then, combining eqs.~(\ref{eq:3}) and~(\ref{eq:5}), we obtain
\begin{eqnarray}
&&\sigma_0 \left( V(t) +  \frac{{\rm d}R(t)}{{\rm d}t} \right)= \frac{{\rm d}N(t)}{{\rm d}t}
\label{fore_mass_cons} \\
\Leftrightarrow && f^{p_z-p_\nu} R(t)^{-1} \simeq R(t)^{(1-\nu)/\nu} \frac{{\rm d}R(t)}{{\rm d}t} - f^{-p_\nu} \frac{{\rm d}R(t)}{{\rm d}t}
\nonumber \\
\label{eq:9}
\end{eqnarray}
where eqs.~(\ref{eq:2}),~(\ref{eq:4}) and the expression for the line density at the pore (see eq.~(\ref{eq:3})) are used.
The tension propagation law and the propagation time in leading order can be expressed in the form of eqs.~(\ref{eq:7}) and~(\ref{eq:8}), respectively, but now the exponent $\beta$ and $\gamma$ are modified as $\beta= p_z-p_\nu$ and $\gamma=\nu/(1+\nu)$.
The deduced exponents coincide with those based on the iso-flux model~\cite{JPCB_Rowghanian_Grosberg_2010,Arxiv_Dubbeldam_2011} (see below).

\begin{figure}
\begin{center}
\includegraphics{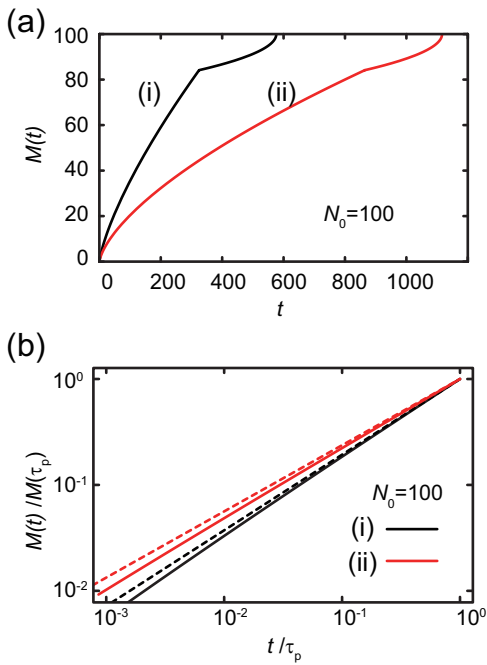}
      \caption{
	  (Color Online) Number $M(t)$ of the translocated segments as a function of time calculated according to the steady-state ansatz (i) or (ii) with $\nu=3/5$, $z=3$ and $N_0=100$.
	  (a) Entire process in normal scale. 
	  (b) Propagation stage in double-logarithmic scale.
	  The dashed lines represent the asymptotic scaling with the slope $\gamma / \nu$.
	  }
\label{fig2}
\end{center}
\end{figure}

The difference in the dynamical evolutions of the number $M(t)$ of the translocated segments depending on the adopted ansatz is plotted in Fig.~\ref{fig2}.
The inflection points in Fig.~\ref{fig2}(a) correspond to the end of the propagation stage, when most of segments have been already sucked. 
The subsequent post-propagation stage can be analyzed using eqs. (\ref{eq:6})/(\ref{fore_mass_cons})~\cite{Saito_EPJE}, where, in contrast to the propagation stage, the growth rate in $M(t)$ increases with time due to the decrease in the overall friction with the process advanced. As shown in Fig. 2(b), the growth of $M(t)$ in the propagation stage follows the asymptotic dynamical scaling $M(t) \sim t^{\gamma/\nu}$, with the slight deviation in the early period due to the finite-size effect.

{\it Iso-flux model}--
We now attempt to resolve the discrepancy between the above two results.
The analysis of eqs.~(\ref{eq:1}),~(\ref{eq:3}) and~(\ref{eq:4}) reveals a unique property of the moving domain in the translocation process.
Given the average relation eq.~(\ref{eq:4}) between $R(t)$ and $N(t)$ , one can show that the last terms in eqs.~(\ref{eq:1}) and~(\ref{eq:3}) are irrelevant, i.e., ${\rm d}N(t)/{\rm d}t > \sigma_R(t) {\rm d}R/{\rm d}t$ and $N(t) > R f^{-p_\nu}$
\footnote{First inequality: Let us write the force balance on the rear end blob $\xi_R^{z-2}v_R \simeq 1/\xi_R$. Since $\xi_R < R$, this leads to the inequality $R^{z-2}v_R >1/R$. Eliminating $v_R$ through $\sigma_R \simeq g_R/\xi_R \simeq v_R^{-(1-\nu)/[(z-1)\nu]}$, one finds $N^{1-\nu}> \sigma_R$, which is equivalent to be proven with eq.~(\ref{eq:4}). 
Second inequality: Using eq.~(\ref{eq:4}), this can be rewritten as $f^{(1-\nu)/\nu}R^{(1-\nu)/\nu}>1$, which holds under the condition $fR>1$ of interest.}.
It then follows that
\begin{eqnarray}
M(t) \simeq N(t)
\label{eq:10}
\end{eqnarray}
and
\begin{eqnarray}
j_R(t) \simeq j_0(t)
\label{eq:11}
\end{eqnarray}
In the scaling level, the fluxes into and out of moving domains balance, as was recently proposed by Rowghanian and Grosberg~\cite{JPCB_Rowghanian_Grosberg_2010}.
To check the validity of the steady-state approximation, let us set $V(t) = v(t,x)$ with $\exists x \in [-R(t), 0]$ so that $v_R(t) = V(t) - \delta v_R(t)$ and $v_0(t) = V(t) + \delta v_0(t)$.
The iso-flux condition eq.~(\ref{eq:11}) is written as 
\begin{eqnarray}
&&\sigma_0 [ V(t) + \delta v_0(t) ] \simeq \sigma_R [V(t)-\delta v_R(t)] \nonumber \\
\Leftrightarrow
&&1+\frac{\delta v_0}{V} \simeq W(t)^q \left( 1- \frac{\delta v_R}{V} \right)^{1-q} 
\end{eqnarray}
where $\sigma_R(t)/\sigma_0 \simeq (1-\delta v_R/V)^{-q} W(t)^q$ with $W(t) = fR(t)$ for $f_\sharp \le f \le f^*$ and 
$W(t) = f^*R^*(t)$ for $f^* \le f$~\footnote{The size of flower part $R^*$ satisfies $V(t)R^*(t) \simeq (f^*)^{p_z}$ (eq.~(\ref{eq:2})) in the stem-flower regime.}.
Given the longest relaxation time $\tau_{\rm relax} \simeq R/V$ of the steadily driven domain and the typical shear rate ${\dot \gamma} \simeq \delta v/R = (v_0-v_R)/R$, the steady-state approximation implicitly assumes the condition $\tau_{\rm relax} \ {\dot \gamma} \le 1 \Leftrightarrow  \delta v(t) \le V(t) $. This self-consistency condition is satisfied for the ansatz (ii), i.e., $\delta v_0 =0 \Rightarrow \delta v_R = [1-W(t)^{-q/(1-q)}] V <V$, but not for the ansatz (i), i.e., $\delta v_R =0 \Rightarrow \delta v_0 = [W(t)^q -1] V$ in the force range $f_\sharp \le f \le f^{**}$ where $W(t)^q >1$.

To summarize, the unique driving mode inherent in the translocation process is manifested in the iso-flux condition, which is to be contrasted with the mechanical driving by pulling chain end~\cite{Saito_process,Sakaue_Wada,PRE_Rowghanian_Grosberg_2012}.
In the latter, the moving domain is constituted of all the $N(t)$ segments, so that eqs.~(\ref{eq:3}) and~(\ref{eq:4}) are modified as $N(t) \simeq R f^{-p_\nu}$ and $N(t)^{\nu} \ll R(t)$, respectively. 
It has turned out that our previous ansatz and corresponding predictions~\cite{Saito_EPJE} are incompatible with the iso-flux condition.
From purely theoretical point of view~\cite{Saito_process,JPCB_Rowghanian_Grosberg_2010,Arxiv_Dubbeldam_2011}, the scaling of the propagation time (thus, the translocation time) is given by
\begin{eqnarray}
\tau_{\rm p} &\simeq& N_0^{1+\nu} f^{-(p_z -p_\nu)} \nonumber \\
&=&\left\{
           \begin{array}{ll}
              N_0^{1+\nu}f^{1+(1/\nu)-z} &   \qquad( N_0^{-\nu} \lsim f \lsim 1 ) \\
              N_0^{1+\nu}f^{-1} &  \qquad (1 \lsim f ) 
           \end{array}
        \right.
\end{eqnarray}
It is somewhat puzzling that the prediction based on the ansatz~(i) seem to be better correlated with numerical and real experiments~\cite{Saito_EPJE}. This might be ascribed to finite-size effects and various crossovers detailed in ref.~\cite{Saito_EPJE}; (i) The last term in eq.~(\ref{eq:9}) serves as a correction term to the tension-propagation scaling law (see the inequality above eq.~(\ref{eq:10})). (ii) The post-propagation stage adds a correction term to the translocation time. (iii) Iso-flux condition (eqs.~(\ref{eq:10}) and~(\ref{eq:11})) is also influenced by the finite-size correction. These effects may be enhanced in realistic situations due to the wall and/or the {\it trans}-side effects, which would be dependent on the size and properties of the pore etc. Since the pore flux would be sensitive to such specific pore effects, one may then be tempted to remove the pore flux (eq.~(\ref{eq:5})) from eq.~(\ref{eq:1}) and to follow the route of the ansatz~(i). If the outgoing flux $j_0(t)$ is suppressed enough compared to that expected from the iso-flux condition, the description based on the steady-state ansatz~(i) may attain its apparent suitability.  A very recent theoretical study based on a more ``accurate" two phase description seems to support such a view~\cite{Ikonen_Sung_2011}, where it has been shown that the frictional interaction between the pore and the polymer significantly affects the dynamics for polymers with experimentally relevant, finite chain length.

\section*{Acknowledgements}


This work was supported in part by the JSPS Core-to-Core Program
``International research network for non-equilibrium dynamics of soft matter",
and 
Kyushu University Interdisciplinary Programs in Education and Projects in Research Development.

%
%

\end{document}